\documentclass[aps,pra,showpacs,twoside,twocolumn,10pt]{revtex4-1}
\usepackage[colorlinks=true, citecolor=blue, urlcolor=blue ]{hyperref}
\usepackage{times,epsfig,amssymb,amsfonts,amsmath,bm,subfigure,mathtools,amsthm,braket,soul,enumitem,xcolor,physics,graphics,graphicx,tikz}
\UseRawInputEncoding
\usepackage[normalem]{ulem}
\usepackage{comment}
\usepackage{epstopdf}
\usepackage{relsize}

\begin{document}

\title{
Enhancing work-extraction in quantum batteries via correlated reservoirs}

\author{Sejal Ahuja, Tanoy Kanti Konar, Aditi Sen (De) }

\affiliation{ Harish-Chandra Research Institute,  A CI of Homi Bhabha National Institute, Chhatnag Road, Jhunsi, Prayagraj - $211019$, India}

\begin{abstract}

Going beyond isolated system dynamics, we examine how local and spatially correlated reservoirs influence the work extraction in quantum batteries. By employing a one-dimensional spin-\(1/2\) model coupled to baths via dephasing and amplitude-damping noise, we demonstrate that correlations in reservoirs can significantly enhance battery's performance compared to local noise. In the dephasing scenario, we prove that correlated reservoirs produce a finite amount of extractable work, or ergotropy, during the transient regime when a two-cell battery is initialized in a product state while local noise yields vanishing ergotropy at all times, despite nonvanishing stored energy in both cases. Numerical simulations confirm that this advantage persists across larger system sizes and for both entangled and product initial states. We also find that the dynamics of quantum coherence closely mirror those of ergotropy, highlighting coherence as a key resource underlying the enhanced performance of quantum batteries. Further, we observe that the fraction of stored energy extracted from quantum batteries displays a sharper contrast between the correlated and local reservoirs. Moreover, for dephasing noise, this fraction remains independent of system size, whereas in the amplitude damping case, it exhibits a clear system-size dependence within the transient regime, highlighting distinct operational behaviors under different noise models. In addition, we reveal that when the battery dynamics is governed by an effective Hamiltonian with long-range interactions, it yields higher ergotropy compared to short-range interactions, emphasizing the advantages of reservoir engineering for efficient device design.

\end{abstract}

\maketitle
\section{Introduction}
\label{sec:intro}

Quantum batteries (QBs), emerged as a promising energy storage technology, exploit quantum mechanical principles to surpass the limitations of classical systems. Since the concept was first introduced by Alicki and Fannes \cite{Alicki}, a wide spectrum of research has explored QBs, ranging from theoretical investigations based on quantum information measures \cite{Bera2020QB,ksen_battery_1,mohan_2021_pra,sarkar2025} to diverse QB architectures designed using various quantum systems \cite{andolina2017,andolina2019,santos2019,andolina2020,Modispinchain,srijon2020,srijon21,srijon2021,alba_1_20,alba_1_22,konar_battery_3,ksen_battery_4,arjmandi_pra_2022,santos_pra_2023,ksen_battery_5,Chaki2024Sep,AI_quantum_battery,mesure_battery,remote_charging_battery,topological_quantumbattery,NV_battery,perciavalle2024,battery_ico,arjmandi_pre_2023,Gyhm2024Jan,liu2025}. Importantly, several prototypes have been experimentally realized in a variety of physical platforms, including quantum dots \cite{wennigerexpqdots}, superconducting transmons \cite{superconducting_battrey_1,superconductQBexp,GemmeexpIBMsupercond}, organic semiconductors \cite{Quach2022Jan}, and nuclear magnetic resonance setups \cite{MaheshexpNMR} (see also \cite{battery_rmp_review}).

Quantum devices are inherently impossible to construct in complete isolation, and their unavoidable interaction with surrounding environments poses one of the central challenges in their realization \cite{breuer2002,Guo2023,mondal_pra_2025}. Such environmental coupling typically leads to  the decay of quantum correlations, thereby undermining the advantage in several quantum information processing tasks, for instance, lowering the gate \cite{fogarty_pra_2015} and quantum teleportation fidelity~\cite{badzia_pra_2000,hu_pra_2010}, reducing channel capacity in dense coding~\cite{adami_pra_1997, Shadman_2010, ShadmanKampermannrev}, and negatively affecting other quantum operations \cite{Guo2023}. In order to encounter these detrimental effects, various error mitigation and error correction techniques have been developed \cite{nielsenchuang2000,Maciejewski2020, Brien23}. In parallel, a natural question has also been raised: ``\emph{Can the coupling of an environment or reservoir to quantum devices positively impact their performance, potentially enhancing efficiency instead of merely causing degradation?}" This has been explored from two complementary perspectives. On the one hand, certain types of environment, particularly non-Markovian noise, have been demonstrated to preserve or even raise the performance of devices such as quantum batteries ~\cite{farina_prb_2019,srijon21,ksen_battery_2,tirone_prl_2023,liu_pra_2024,zhang_pra_2025,tirone_pra_2025,song_prl_2025,ghosh2025,yao_pra_2025}. On the other hand, recent advances in reservoir engineering show that dissipation itself can be harnessed to generate valuable quantum resources, such as entanglement, within quantum systems~\cite{gramajo_pra_2021,oliveira_pra_2023,Francesco_2024,zhu_pra_2025}. In particular, correlated reservoirs  have been demonstrated to entangle two or three qubits~\cite{zou_prb_2022,Zou2024,driessen2025, Longstaff_quantum_2023,li_prb_2025},thus enhancing the efficiency of quantum devices and highlighting the counter-intuitive, and favorable role of dissipation.

\begin{figure}
    \centering
    \includegraphics[width=\linewidth]{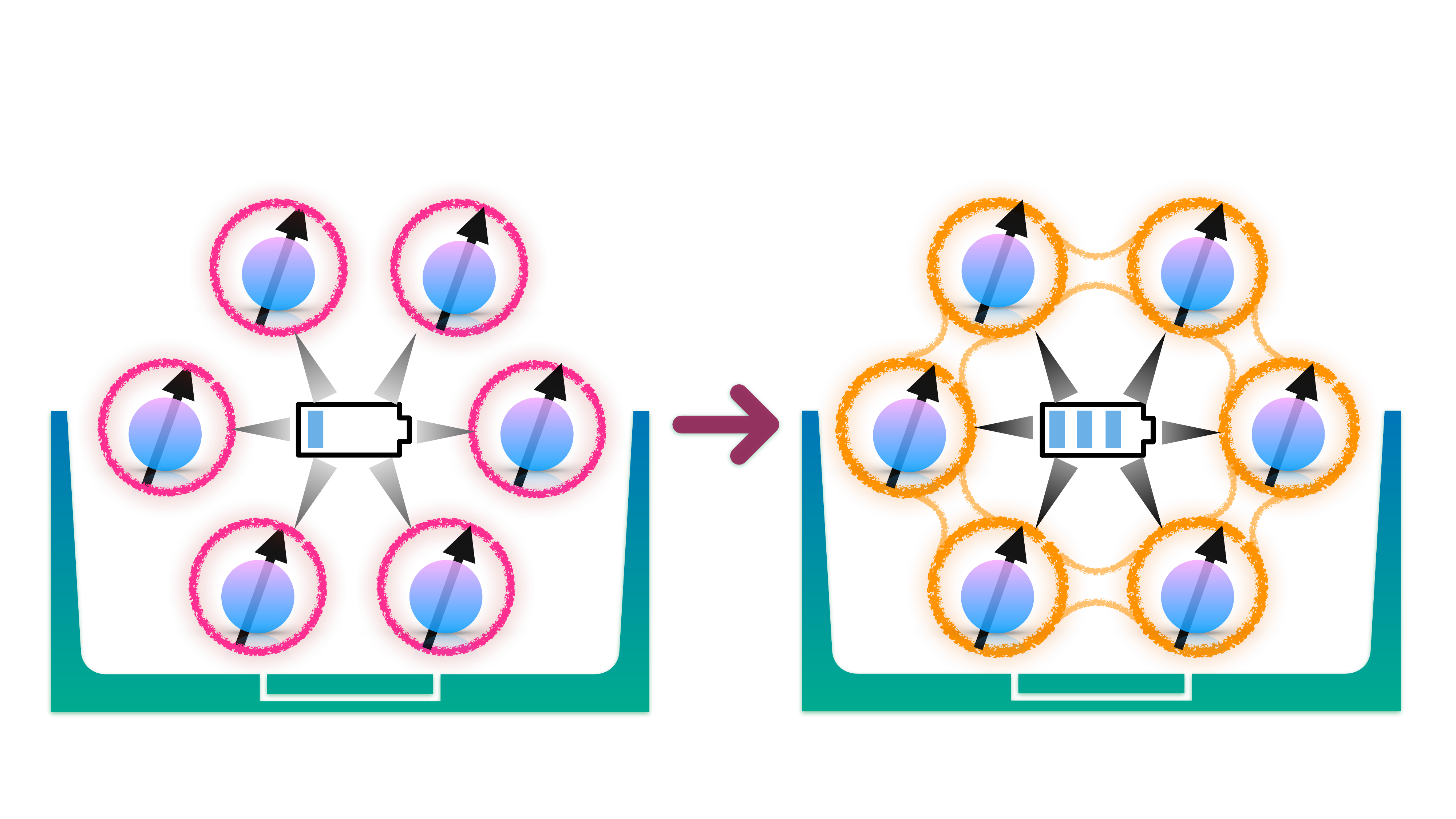}
    \caption{Schematic diagram of charging a quantum battery through correlated noise. Two nearest-neighbor battery cells are coupled to a global environment, which induces an effective interaction between nearest neighbor reservoirs as shown in the right, represented by the curved lines. Initially, the battery is prepared in the ground state of the battery Hamiltonian which can be in a product or an entangled state.  It is then subjected to noise, leading to its charging and evolution towards a final steady state. We call the process correlated noise-induced charging which occurs due to constructive impact of the reservoir. }
    \label{fig:schematics}
\end{figure}

Since the precise resources required to secure quantum advantages in quantum batteries  remain unclear (cf. \cite{Francica2017, andolina2017, squeezingbattery,resourcebattery22, Pawelbattery23}), existing results cannot be straightforwardly applied to guarantee their improved performance through the assistance of reservoirs. To address this gap, we investigate the behavior of a QB coupled to externally correlated reservoirs, an aspect that remains relatively underexplored. Specifically, we ask whether such correlations can enhance the battery's extractable work, namely ergotropy, compared to uncorrelated (local) noise (see Fig. \ref{fig:schematics}). In this framework, we analyze two noise models: correlated dephasing and amplitude damping, which give rise to effective Ising and Dzyaloshinskii-Moriya (DM) type interactions, respectively, between battery cells~\cite{zou_prb_2022,driessen2025}. We prove that two initially uncorrelated battery cells interacting locally with independent reservoirs cannot store any ergotropy under time evolution. Remarkably, introducing correlations between the baths enables energy storage even in the absence of initial correlations among the battery cells, at least in the transient regime. Furthermore, we find that quantum coherence generated in the system at short times serves as the key resource for energy storage, with ergotropy and coherence displaying a one-to-one correspondence when the latter is evaluated in the energy basis, particularly under correlated dephasing noise. Moreover, when initial correlations among cells are incorporated, the ergotropy saturates to a finite value, in stark contrast to the uncorrelated case where it vanishes. These findings remain robust across small to moderate system sizes and for both noise models considered.

We also analyze the fraction of extractable energy and observe distinct behaviors for the two correlated reservoirs considered in this work. In the dephasing case, the fraction remains constant with increasing system-size but decays to zero at long times for initial uncorrelated battery cells. In contrast, under amplitude damping, it exhibits a dependence on the number of cells in the battery and ultimately saturates at a finite value for all system sizes, highlighting a distinct benefit of amplitude damping over dephasing. However, in the case of the correlated initial state, both noise models produce finite fractions of extractable work, which are significantly higher for small time than those obtained via the local reservoirs.  Furthermore, by incorporating effective long-range interactions in dynamics, which are both realistic and ubiquitous in physical systems, we exhibit that the ergotropy of the QB significantly surpasses that obtained under short-range interactions. Altogether, our study demonstrates that noise and correlations are not merely detrimental but can play an essential role in the operation of quantum batteries. By leveraging reservoir correlations and their induced interactions with the battery, it becomes possible to achieve more efficient energy extraction.

The paper is organized as follows. Sec. \ref{sec:setup} introduces the set-up to obtain the dynamics of the battery in the presence of correlated dephasing noise. In Sec. \ref{sec:deph}, we prove the enhancement in ergotropy through correlated reservoirs over local baths and  present the numerical results for different system sizes. In Sec. \ref{sec:adc}, we consider the correlated amplitude damping noise and demonstrate the corresponding results. Sec. \ref{sec:LRint} represents the evolution of the battery under effective long-range interaction  and compares its work-extraction with the short-range ones. Finally, we summarize the results in \ref{sec:conclu}.

\section{The set-up for correlated noise and figures of merits of the battery}
\label{sec:setup}

The main goal of this work is to   demonstrate the role of correlated reservoirs in the performance of quantum batteries.  To do so, we briefly sketch the formalism used to investigate the dynamics of the battery attached to the correlated and local reservoirs. 

\subsection {Quantum battery in contact with correlated environment}

Let us begin by outlining the complete set-up of the battery as a system and its environment. In particular, a quantum battery is modeled by a one-dimensional periodic model consisting of \(N\) spin-\(1/2\) cells. Each cell is coupled to a thermal bath which is either local or spatially correlated (for schematics, see Fig.~\ref{fig:schematics}). One of our goals is to highlight the constructive impact of correlated noise on the storage (extracting) capability of the battery over local noise. The total Hamiltonian of the combined system can be written as  
\begin{eqnarray}
H = H_{B} + H_{E} + H_{BE},
\end{eqnarray}
where \(H_{B}\), \(H_{E}\) and  \(H_{BE}\) represent the battery Hamiltonian,  the Hamiltonian for the environment, and  the system-environment interaction. 

Let us first examine the structure of the environment-Hamiltonian which plays a crucial role in understanding correlated noise, represented as \( H_{E} = \sum_{\textbf{k}} \hbar \omega_{\textbf{k}} b_{\textbf{k}}^{\dagger} b_{\textbf{k}}, \) where \(b_{\textbf{k}}\) (\(b_{\textbf{k}}^{\dagger}\)) denotes the annihilation (creation) of bosonic operator corresponding to the mode \(\textbf{k}\), and \(\omega_{\textbf{k}}\) is the frequency associated with each mode. The initial state of the environment is assumed to be a thermal state, given by   \( \rho_E = \frac{\exp(-\beta H_E)}{\Tr[\exp(-\beta H_E)]}, \) with \(\beta = 1/T\), where \(T\) is the temperature of the environment. The system-bath interactions specifying the influence of the environment on the system are given by  \cite{Zou2024,driessen2025,zou_prb_2022}
\begin{equation}
    H_{BE}^{D}=\sum_{j=1}^{N} \hat{\sigma}_{j}^z\otimes G_j; 
    \quad 
    H_{BE}^{A}=\sum_{j=1}^{N} \hat{\sigma}_{j}^+\otimes G_j+\hat{\sigma}_{j}^-\otimes G_j^\dagger,
    \label{eq:system_bath_coupling}
\end{equation}
where \(H_{BE}^{D}\) represents the dephasing channel, \(H_{BE}^{A}\) denotes the amplitude damping bath, \(G_j=\nu_{\textbf{k}}e^{i\textbf{k}x_j}b_{\textbf{k}}+\text{h.c}\) with \(x_j\) being the position of the \(j^{th}\) battery cell, \(\nu_{\textbf{k}}\) is the coupling strength between the system and the environment, and \(\hat\sigma^{\pm} = \frac{1}{2} (\hat\sigma^x \pm i\hat\sigma^y) \) are the spin ladder operators with \(\hat\sigma^l\) (\(l =x, y, z\)) being the Pauli operators. Under the influence of the environment, the system evolves according to the Gorini-Kossakowski-Sudarshan-Lindblad (GKSL) master equation \cite{breuer2002} which clearly demonstrates three situations, considered in this work -- (1) when all the subsystems (cells) of the battery are attached to the local baths which are non-interacting; (2) when all the cells are attached to the individual bath, and incorporating correlations between the reservoirs leads to the effective nearest-neighbor (NN) interactions, as will be discussed in the succeeding section; (3) the  correlations between the reservoirs are such that the effective long-range interactions between the battery cells emerge.

\subsection{Formalism for correlated dephasing noise}

In order to explore the impact of spatially correlated dephasing noise on the battery dynamics, let us outline the GKSL master equation, describing the dynamics of the reduced density matrix \(\rho_B(t)\) of the system, after tracing out the environment-part from the entire system-environment state \(\rho_{tot}(t)\). Here, the initial state of the battery and the environment are chosen to be the ground state of the battery Hamiltonian \(H_B\), and \(\rho_E\) respectively, thereby the initial state of the entire system being taken as \(\rho_B(0)\otimes \rho_E(0) \) and the battery-reservoir interaction is governed by \(H_{BE}^{D}\).  
Under the assumption of weak system-environment coupling, i.e., by considering small \(\nu_{\textbf{k}}\) compared to the energy splitting of the local system and by applying the Born-Markov approximation, the battery evolves according to the  GKSL master equation given by \cite{breuer2002}
\begin{eqnarray}
\nonumber \dot\rho_B(t) &=& -\frac{1}{\hbar^2}\int_{0}^{\infty} d\tau [\text{Tr}_{E}[H_{BE}^D(t), [H_{BE}^D(\tau),\rho_B(t)\otimes\rho_E]]],
\label{eq::master_eqn_petru}
\end{eqnarray}
where the mathematical expression is written in the interaction picture. This expression simplifies as \cite{Zou2024,driessen2025}
\begin{eqnarray}
\nonumber \dot\rho_B (t) &=& -\frac{1}{\hbar^2}\sum_{r,s}\int d\tau [\Gamma_{rs}(t-\tau)(\hat\sigma_{r}^z \hat\sigma_{s}^z \rho_B - \hat\sigma_{s}^z \rho_B \hat\sigma_{r}^z) + \\ && \Gamma_{sr}(\tau-t)(\rho_B \hat\sigma_s^z \hat\sigma_r^z - \hat\sigma_r^z \rho_B \hat\sigma_s^z)],
\end{eqnarray}
where  the two-point correlation function \(\Gamma_{rs} (t-\tau) \equiv \langle G_r (t) G_s(\tau) \rangle\) with \(G_r(t) = e^{iH_Et/\hbar}G_r(0)e^{-iH_Et/\hbar}\) in the interaction picture, \(\langle \mathcal{Q} \rangle \equiv \text{Tr}_E(\rho_E \mathcal{Q})\), and \(r,s\) being the  battery cells. After rearranging the terms, the final equation responsible for the evolution of the battery takes the form
\begin{eqnarray}
{\dot\rho_B(t)} &=& -i [\mathcal{H}_{z} (t), \rho_B(t)] + \sum_{i,j = 1}^{N} \mathcal{L}_{ij}^{z}(t) \rho_B(t),
\label{eq:mastereqn}
\end{eqnarray}
where \(\mathcal{H}_{z} (t)\) arises from the coherent interaction between qubits induced by correlated reservoirs. It reads as
\begin{eqnarray}
\mathcal{H}_{z} (t) &=& \sum_{j<k=1}^{N}\mathcal{J}^{z}(t) \hat{\sigma}_j^z\hat{\sigma}_{k}^z,
\label{eq:corrdeph}
\end{eqnarray}
with \(\mathcal{J}^{z}(t)\) denoting the interaction strength between the pair of spins, given as
\begin{equation}
    \mathcal{J}^z(t) = \frac{1}{2i\hbar^2}\sum_{i\neq j}\int d\tau [\Gamma_{ij}(\tau) - \Gamma_{ij}(-\tau)].
\end{equation}
This interaction strength depends on the two-point time correlation between time \(\tau\) and \(-\tau\) of the environment operators. In general, \(\mathcal{J}^{z}(t)\) is time-dependent; however, since we consider Markovian noise with time-independent bath correlations, we set \(\mathcal{J}^{z}(t)\equiv\mathcal{J}^{z}\) which is independent of time. The term \(\mathcal{L}_{ij}^{z}(t)\) in Eq. (\ref{eq:mastereqn}) describes the environmental dissipation and is expressed as  
\begin{eqnarray}
\mathcal{L}_{ij}^{z}(t) \rho_B&=& \gamma_{ij}^{z}(t) [\hat{\sigma}_j^z\rho_B\hat{\sigma}_{i}^z - \frac {1}{2} \{\hat{\sigma}_i^z\hat{\sigma}_{j}^z, \rho_B\}],
\end{eqnarray}
where \(\gamma_{ij}^{z}\) representing the dephasing rate, is taken to be a constant with time, i.e., \(\gamma_{ij}^z (t) \equiv \gamma_{ij}^z\), having two components, \(\gamma_{ii}^z\) which are real while \(\gamma_{i\neq j}^z\) can be complex. Note here that, in general, \(\gamma_{ij}^z (t) = \frac{1}{\hbar^2} \int_{-t}^{t} d\tau \Gamma_{ij}(\tau)\). Here, \(\gamma_{ii}^{z}\) and \(\gamma_{i\neq j}^{z}\) correspond to the local and the correlated dephasing rate respectively, where imaginary part of \(\gamma_{i\ne j}^{z}\) captures the  quantum correlations present among the reservoirs respectively.

\section{Enhanced ergotropy via correlated noise}
\label{sec:deph}

We now report the benefit of correlated dephasing noise over the local one. To examine the role of correlations, we consider two types of initial state of the battery - (1) the ground state, which is a product state without correlations, and (2) the entangled state. The battery Hamiltonian is considered to be  transverse Ising model \cite{sachdev09},
\begin{eqnarray}
    H_{B} &=& \sum_{i=1}^{N} \frac{h'}{2} \hat\sigma_i^x+\frac{J'}{4}\sum_{i=1}^{N}\hat\sigma_i^z\hat\sigma_{i+1}^z
    \label{eq::battery_hamiltonian}
\end{eqnarray}
where \(h'\) and \(J'\) denote the external magnetic field strength and the coupling constant between battery cells, respectively. In order to perform all the analysis unit-independent, we choose \(h=h'/J'\) (provided \(J' \neq 0\)) and also consider the periodic boundary condition, i.e., \(\hat \sigma_{N+1}\equiv \hat \sigma_1\). 

\emph{Ergotropy.} With this battery Hamiltonian which is subject to the correlated or local dephasing noise, we assess the performance of the battery in terms of the maximum amount of extractable work, i.e., ergotropy~\cite{Alicki, Batteryreview}. It is defined as
\begin{eqnarray}
\mathcal{E}(t) = \text{Tr}[H_{B}\rho_B(t)] - \min_{U} \text{Tr}[H_{B}U\rho_B(t)U^{\dagger}],
\end{eqnarray}
where minimization is carried out over all unitary operators, \(\rho_B(t)\) denotes the evolved state of the battery obtained via the GKSL master equation through Eq. (\ref{eq:master_eq}) and the energy of the resulting state is given as \(W(t)=\text{Tr}[H_{B}\rho_B(t)]\). By using spectral decomposition, ergotropy reduces to \(\mathcal{E}(t) = \sum \lambda_i \epsilon_i\) where \(\epsilon_i\) and \(\lambda_i\) are the eigenspectrum of \(H_B\) and \(\rho_B(t)\) respectively with the condition \(\lambda_{i+1}\geq \lambda_i\) and \(\epsilon_{i+1}\leq \epsilon_i\). Here, \(\lambda_i\) can be called the population of each level of the battery system. With the entire set-up in hand, we are now ready to establish the gain in ergotropy obtained with the aid of correlated reservoir attached to battery cells.

In the first case, we specifically consider that the initial state of the battery is the ground state of  \(H_B\) having no interaction term, i.e., \(J'=0\) giving \(H_{B} = \sum_{i=1}^{N} \frac{h'}{2} \hat\sigma_i^x\) and hence \(\rho_B(0) = (|-\rangle\langle-|)^{\otimes N}\). In this case, no charging is applied, but after preparation, the cells of the battery are connected to the correlated or local reservoirs. Note that the distinction between local and correlated noise lies in the ability to generate effective interactions between the battery cells. In particular, under local noise, each spin is influenced independently by the bath, with no possibility of producing interaction between the spins of the battery. In contrast, correlated noise can induce effective interactions between the spins; in particular, the correlated dephasing channel leads to an effective Ising-type interaction between the battery cells, as given in Eq. (\ref{eq:corrdeph}) which finally becomes responsible for achieving nonvanishing ergotropy. We call this positive consequence of the reservoir on the performance of the quantum battery as {\it constructive impact}. We now present the following proposition for two battery cells which can then be confirmed numerically for a moderate number of systems under dephasing correlated reservoirs. 

\textbf{Proposition.} {\it At finite time, correlated dephasing reservoirs can provide nonvanishing ergotropy of the battery which is initially in a product state, whereas local dephasing noise cannot extract any work from the same battery at any time.}  

\begin{proof} 

The evolution of the battery consisting of two cells under local and correlated noise are described respectively by  
\begin{eqnarray}
        \eval{\frac{d\rho_B}{dt}}_{loc}&=& \sum_{i=1}^2\gamma_{ii}^{z} [\hat{\sigma}_i^z\rho_B\hat{\sigma}_{i}^z - \frac {1}{2} \{\hat{\sigma}_i^z\hat{\sigma}_{i}^z, \rho_B\}], \\
        \eval{\frac{d\rho_B}{dt}}_{cor}&=&\nonumber-i[\mathcal{H}_z,\rho_B]+\sum_{i,j=1}^2\gamma_{ij}^{z}[\hat{\sigma}_j^z\rho_B\hat{\sigma}_{i}^z - \frac {1}{2} \{\hat{\sigma}_i^z\hat{\sigma}_{j}^z, \rho_B\}],\\
\end{eqnarray}
where \(\mathcal{H}_z= \mathcal{J}_z\hat\sigma_i^z\otimes \hat\sigma_{i+1}^z\). 
Under local noise, the transformation of the initial state occurs as  
\(\rho_B(0)=\otimes_{i=1}^2 \rho_i(0) \rightarrow \rho_B(t)_{loc}=\otimes_{i=1}^2 \rho_i(t)\), where  
\[
\rho_i(t)=\frac{1}{2}\left [
\begin{array}{cc}
 1 & - e^{-2 \gamma  t} \\
 -e^{-2 \gamma  t} & 1 \\
\end{array}
\right ]; \,\, i=1, 2,
\]
with \(\gamma_{ii}^z\equiv\gamma\). The corresponding energy and the ergotropy of the battery are respectively given by \(W(t)= - h e^{-2 \text{$\gamma $} t}\) and \(\mathcal{E}(t)=0\) for all times.  

On the other hand, for the correlated case, the evolved density matrix becomes  
\begin{widetext}
    \begin{equation}
    \rho_B(t)_{cor}=\frac{1}{4}\left[
\begin{array}{cccc}
  1 & -e^{-i t (-2 i \text{$\gamma $}+\mathcal{J}_z+2 q)} & -e^{-i t (\mathcal{J}_z-2 (q+i \gamma))} & e^{-4 t (\gamma+p)} \\
 - e^{i t (2 i \gamma+\mathcal{J}_z+2 q)} & 1 & e^{4 t (p-\gamma)} & - e^{i t (2 i \gamma+\mathcal{J}_z-2 q)} \\
 - e^{i t (2 i \gamma+\mathcal{J}_z-2 q)} &  e^{4 t (p-\gamma)} & 1 & -e^{i t (2 i \gamma+\mathcal{J}_z+2 q)} \\
  e^{-4 t (\gamma+p)} & -e^{-i t (\mathcal{J}_z-2 (q+i \gamma))} & -e^{-i t (-2 i \gamma+\mathcal{J}_z+2 q)} & 1 \\
\end{array}
\right ],
\end{equation}
\end{widetext}
where \(\gamma_{ii}^z \equiv \gamma\) and \(\gamma_{12}^z=p+iq\), with the imaginary part \(q\) describing the quantum correlated noise. In this case, \(W(t)=-h e^{-2 \text{$\gamma $} t} \cos (2\mathcal{J}_z t) \cos (2 t q)\) and obtaining a closed-form expression for the ergotropy is challenging due to the dependence of eigenvalues on system and environment parameters. However, in the limit \(\gamma\gg |\gamma_{12}^z|\), the ergotropy simplifies to  
\begin{eqnarray}
    \mathcal{E}(t)&=&W(t)+\frac{he^{-4 \gamma t}}{2}  \sqrt{\sinh^{2}(4 t p)  +  4 e^{4 t \gamma} \cos^{2}(2 t q)},\,\,\,\,\,\,
\label{eq:ergodeph}
\end{eqnarray}
which is clearly nonvanishing for a finite time and vanishes for \(t \rightarrow \infty\). Note that we recover the expression for local noise by putting \(p=q=\mathcal{J}_z =0\) in Eq. (\ref{eq:ergodeph}).
Therefore, we identify a situation, at which the ergotropy remains nonvanishing in the transient time under correlated noise which cannot be seen in the local noise case. Hence proved.
\end{proof}

{\it Note 1.} The above analysis reveals that when local reservoirs are individually connected to each battery cell,  each cell becomes excited, leading to an increase in stored energy although extractable work (ergotropy) cannot be produced. This occurs because the ground state taken as the initial state is a completely passive state. Specifically, the ground state is fully occupied, and although local noise can populate the excited states, and thereby raise the system's energy, it cannot invert the population such that the excited states become more populated than the ground state.  More precisely, if the initial population is given by \( \lambda_0(0) = 1 \) for the ground state and \( \lambda_i(0) = 0 \) for the excited states, \( (i \in \{1, 2, 3\} )\), during the evolution under local noise, the populations of the energy levels change. However, the ordering in the populations remains the same, i.e., \( \lambda_0(t) \geq \lambda_1(t) \geq \lambda_2(t) \geq \lambda_3(t)\), which corresponds to a passive state structure, hence the ergotropy remains zero. In contrast, under a correlated noise, an effective interaction is induced between the spins which disturbs the orderings of the equilibrium population distribution among the energy levels, potentially resulting in population inversion or imbalance. As a result, at transient times, the system exhibits nonvanishing ergotropy, indicating the presence of extractable work.

\emph{Note 2.} In the case of correlated dephasing noise,  there is a competition  between the effective interaction \(\mathcal{J}_z\) and the local dephasing strength \(\gamma\) in Eq. (\ref{eq:mastereqn}). In particular,  when \(t< t_c\), the interaction strength is higher than \(\gamma\), leading to a  gain in physical quantities due to the correlation between reservoirs, although for \(t > t_c\), the local dephasing process dominates, driving the state towards being diagonal in the energy basis, i.e., a maximally mixed state, where the ergotropy eventually vanishes.

\subsection{Constructive impact of correlated dephasing noise on multiple cells of quantum battery}

Let us now analyze the ergotropy of a quantum battery containing \(N(>2)\) number of cells, which are in contact with correlated and local dephasing reservoirs. We again compare two situations.

{\it Local baths connected to individual cells:} Initializing the battery into  \(|\psi_B\rangle = |-\rangle^{\otimes N}\) and after the local action of the dephasing noise on individual cells, we obtain \(\rho_B(t)_{loc}=\otimes_{j=1}^{N}\rho(t)\). It cannot generate any classical as well as quantum correlations between spins. Consequently, the ergotropy remains vanishing, implying vanishing extractable work. 

\begin{figure}
    \centering
    \includegraphics[width=\linewidth]{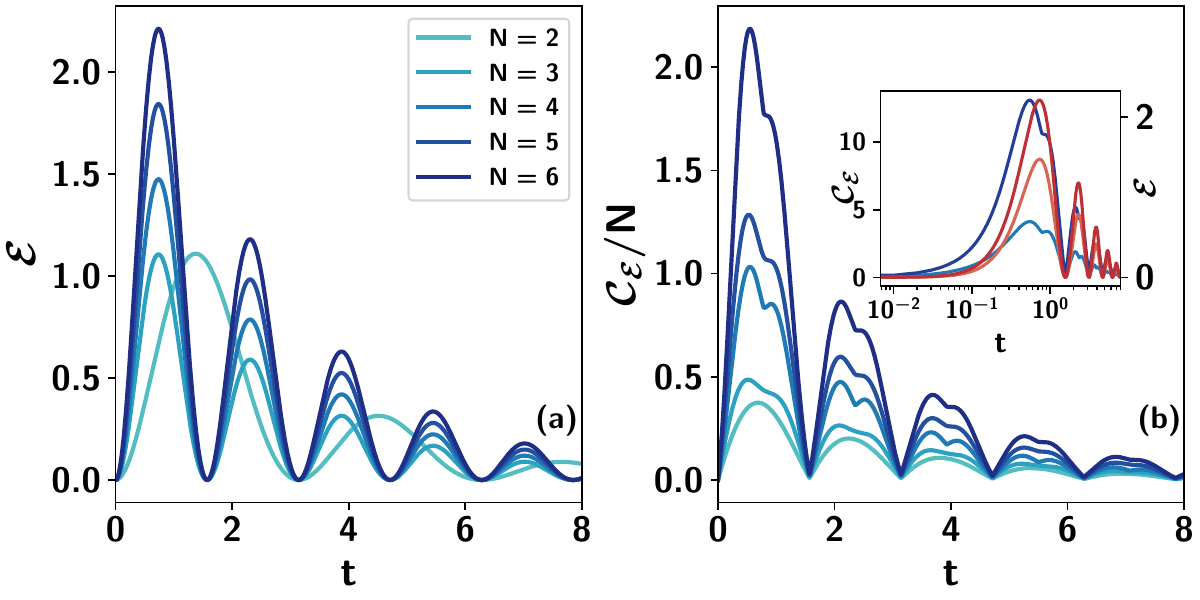}
    \caption{(a) Ergotropy, \(\mathcal{E}\) (ordinate) against time, \(t\) (abscissa) for different system sizes, \(N\). The initial state is taken to be the ground of \(H_B\) with \(J'=0\).  A nonvanishing ergotropy which increases with \(N\) is obtained in the transient time and it decreases to zero at the steady state while it vanishes for all time with local reservoir. (b) Plot of quantum coherence measured via \(l_1\)-norm, \(\mathcal{C}_E/N\) in the energy basis of the battery Hamiltonian of the dynamical state (ordinate) with time (abscissa) for different system sizes. Comparing (a) and (b), it seems that the resource responsible for producing a finite ergotropy with correlated reservoir is quantum coherence  (see also the inset). Other parameters of the system are given as \(\gamma=0.2\), \(\gamma_{12}^z=0.01e^{i\pi/3}\), \(\mathcal{J}_z=1\) and \(h=1\). All axes are dimensionless.}
    \label{fig:dephasing_N}
\end{figure}

{\it Correlated reservoirs:} In this scenario, the effective Hamiltonian \(\mathcal{H}_z\) contains nearest-neighbor interactions, and hence can, in principle, create both quantum and classical correlations \cite{Bera_2018,kmodi_rev} in the system even when the initial state is a product. Hence, it leads to a nonvanishing ergotropy for an arbitrarily large system-size (see Fig. \ref{fig:dephasing_N}) as also shown in Proposition for \(N=2\).
Specifically, the effective interaction between the spins is described by  
\[
\mathcal{H}_z=\mathcal{J}_z\sum_{j=1}^{N}\sigma_j^z\sigma_{j+1}^z.
\]
Correspondingly, the dissipation matrix between the spins is given by  
\begin{equation}
    \hat\Gamma^z = \begin{bmatrix}
\gamma & \gamma_{12}^z & 0 & \cdots & 0 & \gamma_{1N}^{z^*} \\
\gamma_{12}^{z^*} & \gamma & \gamma_{23}^z & \ddots & & 0 \\
0 & \gamma_{23}^{z^*} & \gamma & \ddots & \ddots & \vdots \\
\vdots & \ddots & \ddots & \ddots & \gamma_{N-2N-1}^z & 0 \\
0 & & \ddots & \gamma_{N-2N-1}^{z^*} & \gamma & \gamma_{N-1N}^z \\
\gamma_{1N}^z & 0 & \cdots & 0 & \gamma_{N-1N}^{z^*} & \gamma
\end{bmatrix},
\label{eq::gama_matrix}
\end{equation}
where \(\hat\Gamma^z\) captures all the strength of both the correlated and local noise and we choose \(\gamma_{ii+1} = \gamma_{12}^z\), i.e., all the nearest-neighbor dissipative couplings are equal. In order for the GKSL evolution to represent a completely positive trace-preserving (CPTP) map, the matrix \(\hat\Gamma^z\) must be positive semidefinite, i.e., \(\hat\Gamma^z\ge 0\). For the tridiagonal structure of \(\hat\Gamma^z\), the eigenvalues of the matrix can be computed analytically and are given by  
\begin{equation}
\hat \Gamma^z_m=\gamma+2\abs{\gamma_{12}^z}\cos(k_m+\phi),
\end{equation}
where \(k_m=\frac{2\pi m}{N}\) and \(\phi=\arg(\gamma_{12}^z)\). The CPTP condition ensures that \(\gamma\ge 2\abs{\gamma_{12}^z}\), which must be taken care of while choosing the noise parameters. Our numerical simulations for moderate \(N\) reveals the following (see Fig. \ref{fig:dephasing_N} ): (1) Ergotropy oscillates with time irrespective of \(N\) before it vanishes for large time, \(t\).  It originates from the creation of effective interactions among the spins, leading to a damped oscillatory behavior independent of the system-size. (2) In the transient time, the amplitude of \(\mathcal{E}\) decreases with time. It indicates that beyond a certain time \(t>t_c\), the dissipation term dominates, driving the battery state toward a fully dephased state in the energy basis. (3) The maximum ergotropy obtained in the first oscillation increases with increasing number of cells, \(N\).

{\it Resource responsible for constructive effect on battery.} These observations lead to an immediate question - ``What is the resource produced due to correlated noise that is responsible for a finite ergotropy?". We answer this query by identifying that the quantum coherence in the energy basis acts as a resource in the battery. In particular, we compute \(l_1\)-norm of coherence~\cite{coherence_review} defined as
\begin{equation}
    \mathcal{C_E}=\sum_{i\ne j}|\rho'_{ij}|
\end{equation}
where \(\rho'=\sum_{n,n'}\ket{E_n}\bra{E_n}\rho(t)\ket{E_{n'}}\bra{E_{n'}}\) is written in the basis of the battery Hamiltonian, \(H_B\). Interestingly, we observe that the coherence behaves almost in a similar fashion as \(\mathcal{E}\) with time. In particular, when coherence vanishes, the ergotropy also becomes zero. One can expect this as the effective interactions among the spins generate coherence between the energy levels of the battery, leading to population imbalances in the system that manifest as ergotropy. However, as evolution proceeds, local noise gradually destroys this coherence, causing the ergotropy to decay with time.  

\begin{figure}
    \centering
    \includegraphics[width=\linewidth]{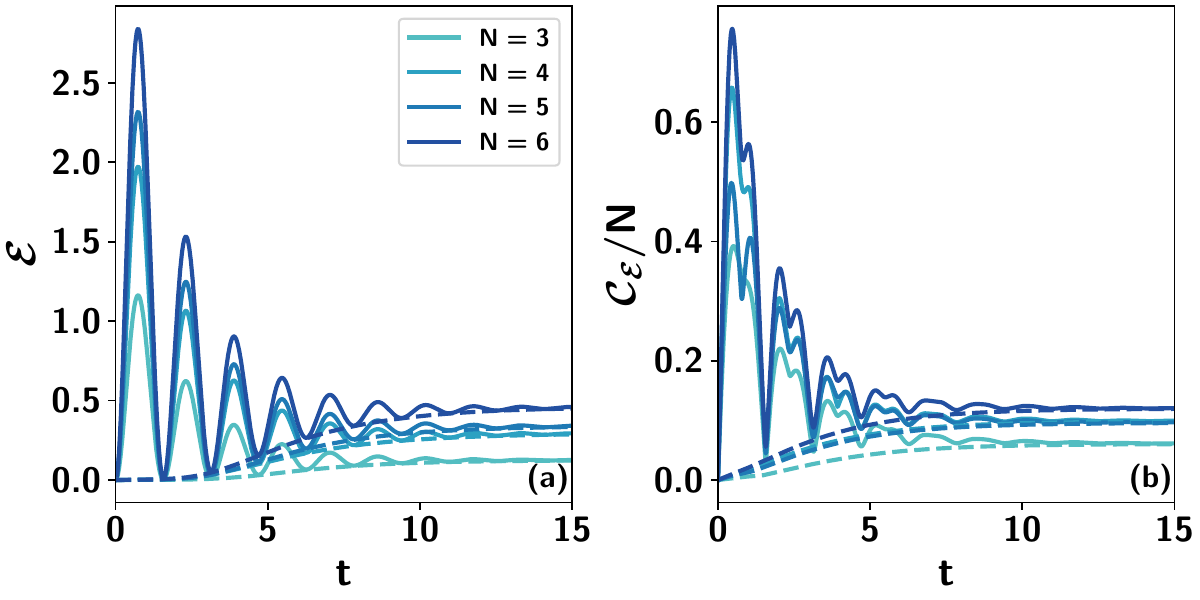}
    \caption{(a) Ergotropy and (b) coherence (ordinate) vs time (abscissa). The initial state is the ground state of battery Hamiltonian \(H_B\) with \(h=1.3\) under correlated (solid lines) and local (dashed lines) dephasing reservoirs. Unlike  Fig. \ref{fig:dephasing_N}, both the baths produce nonvanishing ergotropy. Clearly, correlated reservoirs can produce higher ergotropy and coherence in the transient time compared to the local noise. However, at the steady state, both converge to a same value.   
    All other specifications are same as in Fig. \ref{fig:dephasing_N}. All axes are dimensionless. }
    \label{fig:corrlated_ndep}
\end{figure}

\subsection{ Effect of correlated noise on initially entangled battery} 

In order to showcase the initial correlation in the battery, the initial state is taken to be the ground state of the transverse Ising model, in Eq. (\ref{eq::battery_hamiltonian}) where both \(h'\) and \(J'\) are comparable, so that the state can, in principle, be entangled. Unlike the product state,  the ergotropy turns out to be nonvanishing, both in the transient and steady state regimes, in the presence of both local and correlated noise. Specifically, unlike the previous situations, the impact of local and correlated noise on the work extraction capability of the battery is not so drastic, although these are some notable differences, again for a small time period. Let us illustrate these features, especially the advantages of the correlated noise over the local ones.

In the presence of initial entanglement, local dephasing noise can induce population imbalances between the energy levels, which are then stored as the ergotropy of the battery (see dashed lines in Fig.~\ref{fig:corrlated_ndep}(a)). Moreover, the system-size plays a significant role: as the system-size increases, the ergotropy exhibits a non-monotonic growth, which may arise due to the finite-size effects.  In contrast, the presence of correlated noise  significantly alters the picture. In this case, there is a distinct benefit in  ergotropy during the transient regime, irrespective of system-size. Specifically, the  large amount of work extraction capabilities from the battery emerges in the presence of correlated reservoir although
 it saturates to the same value as in the local noise case for large time. As argued before, this indicates that after a critical time, the local noise dominates, and the influence of correlated noise vanishes (see Fig.~\ref{fig:corrlated_ndep}(a)). Similar to the local noise case, the ergotropy during the transient regime exhibits a non-monotonic dependence on the system-size.

Again, we find that coherence in the energy basis acts as the key resource for the battery, exhibiting a similar behavior to that of ergotropy: under correlated noise, coherence generation is significantly enhanced in the transient time which is reduced for \(t> t_c\) when the local noise part, i.e., \(\gamma_{ii}\) dominates with vanishing \(\gamma_{ij}\)(s) (see Fig.~\ref{fig:corrlated_ndep}(b)). Interestingly, we observe that the coherence in the transient domain is higher for product states than the one obtained with entangled initial states. This can be attributed to the fact that starting from a product state, the generation of coherence is large, which enables the storage of nonvanishing ergotropy. 
\begin{figure}
    \centering
    \includegraphics[width=\linewidth]{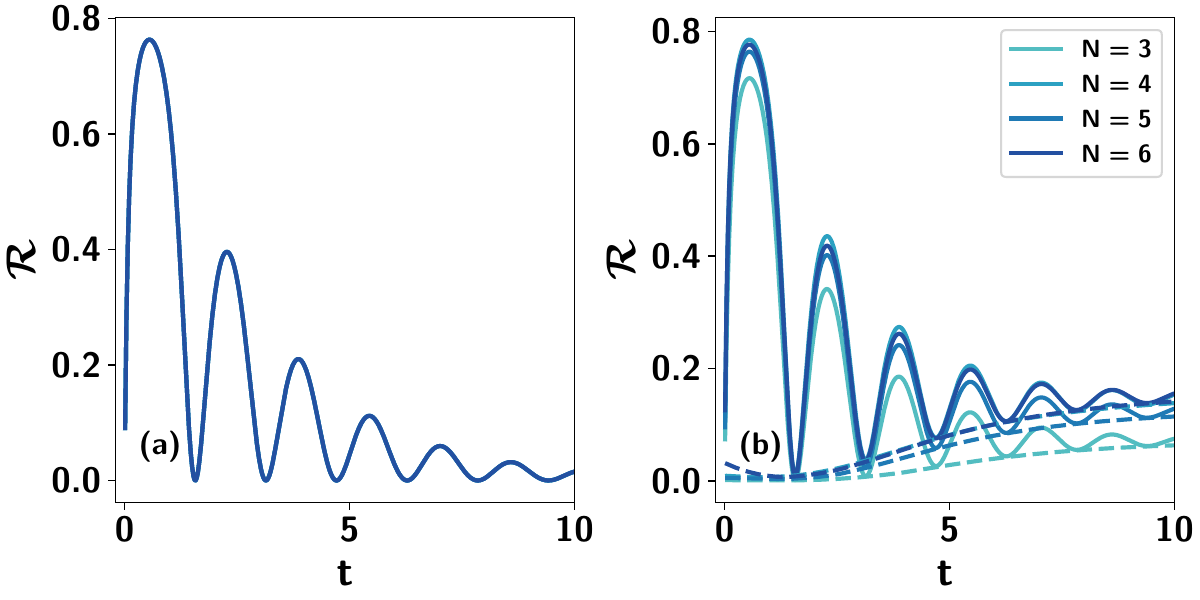}
    \caption{The ratio of ergotropy and stored energy, \(\mathcal{R}\) (ordinate) with time, \(t\) (abscissa) by varying \(N\) under correlated and local reservoirs. (a) The initial state is in a product state with the battery Hamiltonian, \(J = 0\) and (b) corresponds to the ground state of \(H_B\) with \(h=1.3\). Interestingly, \(\mathcal{R}\) is independent of system-size for product state while it depends on \(N\), especially for small \(N\).  The initial state dependence is also evident by comparing \(\mathcal{R}\) in (a) and (b). In particular, it saturates to a non-zero value for entangled initial states in contrast to the product state where it vanishes. All other system parameters are same as in Fig. \ref{fig:dephasing_N}. All axes are dimensionless.}
    \label{fig:dephase_ratio}
\end{figure}

\subsection{Fraction of extractable energy: correlated vs local environments}

It is instructive to study the fraction of stored energy that can be extracted as useful work~\cite{farina_prb_2019}, defined as  
\begin{eqnarray}
    \mathcal{R} \equiv \frac{\mathcal{E}(t)}{E(t)},
\end{eqnarray}
where \(\mathcal{E}(t)\) and \(E(t)\) denote the ergotropy and total energy stored in the battery at time \(t\) which is defined as \(E(t)=W(t)-W(0)\) respectively. The ratio \(\mathcal{R}\) quantifies the efficiency of extracting work from the stored energy, since, in general, a portion of  stored energy may not be extractable due to dissipation losses. One observes that when the initial state is a product state, the local dephasing channel cannot alter the populations of the energy levels, leading to \(\mathcal{R}=0\) throughout the entire evolution. This highlights the disadvantage of local dephasing noise. In contrast, if the battery is charged under correlated dephasing noise starting from a product state, a non-zero \(\mathcal{R}\) emerges during the transient time, demonstrating the advantage of correlated noise. However, for large time, i.e., in the steady state, it again vanishes (see Fig.~\ref{fig:dephase_ratio}). Interestingly, our results reveal that the fraction of extractable energy is independent of the system-size \(N\), indicating that, on average, the same fraction of work can be extracted regardless of the number of qubits. This further underscores the beneficial role of correlated dephasing noise compared to its local counterpart.

Let us now analyze \(\mathcal{R}\) both for local and correlated reservoirs,  when the initial state is entangled.  Again, from the behavior of \(\mathcal{R}\), a clear distinction can be found between local and correlated reservoirs although some contrasting behaviors emerge with the product and entangled initial states  (comparing Figs. \ref{fig:dephase_ratio}(a) and (b)). In particular,  we observe that \(\mathcal{R}\) depends on the system-size although with high \(N\), it becomes almost system-size independent.  
In the transient regime, the overall behavior of \(\mathcal{R}\) is qualitatively similar for both product and entangled initial states. However, at longer times, only the entangled initial state exhibits convergence to a finite value, irrespective of whether the noise is local or correlated. This indicates that steady-state behavior originates from the initial entanglement of the battery, whereas transient time performance is governed primarily by the spatial correlations introduced through the reservoirs, highlighting a clear separation of contributing factors.


\section{Quantum Battery set-up with amplitude damping correlation of reservoirs}
\label{sec:adc}

After establishing the benefit of correlated dephasing noise, we address the following question - "Does the benefit of correlated noise persist, even when the nature of noise changes?". We indeed exhibit that when the noise is of amplitude damping kind and there is a correlation between the reservoirs, the extractable work of the battery can also be induced. Further, we compare the outcomes with those obtained under the dephasing noise, and determine which of the two noise types offer better battery performance.

{\it GKSL master equation for amplitude damping correlated baths.}  
In this case, the battery and the reservoir Hamiltonian remain the same, although the battery-reservoir interaction changes, i.e., \(H_{BE}^{A}=\sum_{j=1}^{N} \hat{\sigma}_{j}^+\otimes G_j+\hat{\sigma}_{j}^-\otimes G_j^\dagger\), which again
 couples two nearest neighbor reservoirs and they interact individually to nearest-neighbor battery cells. In the weak-coupling limit, we can determine the quantum master equation that governs the system's evolution (see Appendix \ref{sec:appendixB} for temperature-dependence of the reservoir) \cite{}. 
Hence, the GKSL master equation controlling the dynamics of the battery reads as 
\begin{eqnarray}
{\dot\rho_B(t)} &=& -i[\mathcal{H}_{xy}, \rho_B(t)] + \sum_{i,j=1}^N \mathcal{L}_{ij}(t) \rho_B(t),
\label{eq:mastereqn_adc}
\end{eqnarray}
where \(\mathcal{H}_{xy}\) is the effective interaction between the battery cells, given by 
\begin{eqnarray}
{\mathcal{H}_{xy}} &=& \sum_{i}[\mathcal{J} \hat \sigma_{i}^{+} \hat \sigma_{i+1}^{-} + \mathcal{J^{*}} \hat \sigma_{i}^{-} \hat \sigma_{i+1}^{+} ]  \nonumber \\
&=&\sum_{i=1}^{N} [\frac{J}{2} (\hat\sigma_i^x\hat{\sigma}_{i+1}^x + \hat\sigma_i^y\hat\sigma_{i+1}^y) \nonumber + \frac{D}{2}(\hat\sigma_i^x\hat\sigma_{i+1}^y - \hat\sigma_i^y\hat\sigma_{i+1}^x)],\\
\label{eq:adccorr}
\end{eqnarray}
where \(J\) and \(D\) are the \(XX\) and Dzyaloshinskii-Moriya interaction strengths between the sites generated due to the correlation between the baths with \(\mathcal{J}=J+iD\).  Again, it is considered to be time-independent in the Markovian limit. On the other hand, the Lindbladian, \(\mathcal{L}_{ij}(t)\) representing the dissipation term can be written as
\begin{eqnarray}
\mathcal{L}_{ij}(t) \rho_B(t)&=& \gamma_{ij}(t) [\hat \sigma^-_{j}\rho_B\hat \sigma^+_{i} - \frac {1}{2} \{\hat \sigma^+_{i} \hat \sigma^-_j, \rho_B\} ],
\end{eqnarray}
where 
\(\gamma_{ij}\)s are the dissipation strength as in Eq. (\ref{eq::gama_matrix}).  Notice further that although  Lindbladian of this noise model differs from the previous case, the properties of the correlation matrix, \(\hat{\Gamma}^z\) in both situations remain same. We now study the pattern of the ergotropy for the quantum battery with time. 

\begin{figure}
    \centering
    \includegraphics[width=\linewidth]{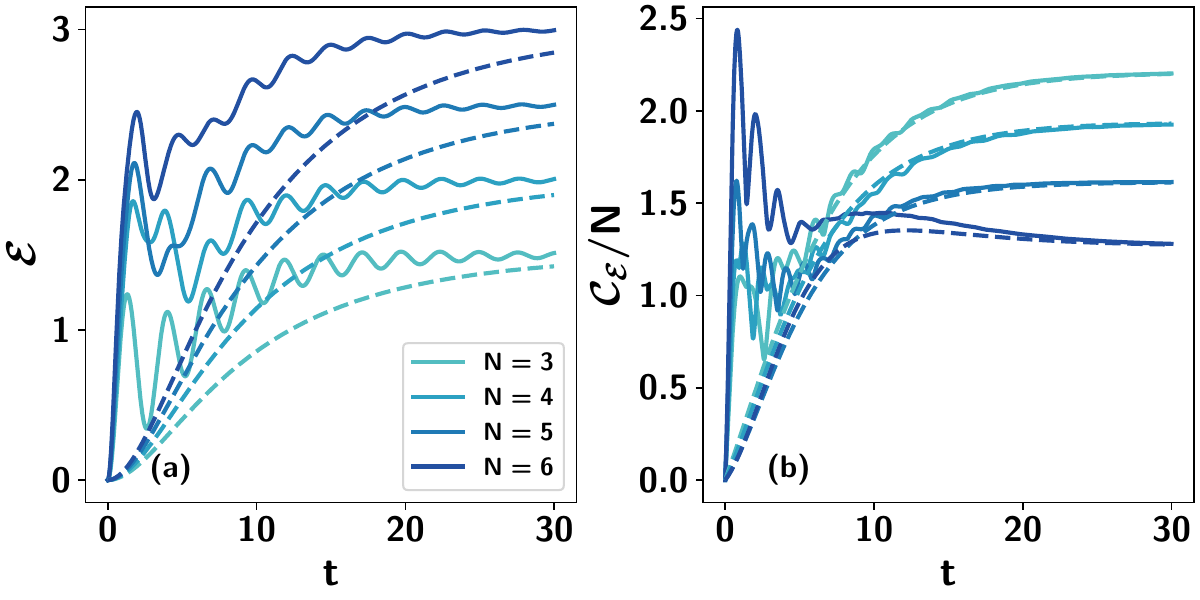}
    \caption{(a) The ergotropy (vertical axis)  with respect to time (horizontal axis) for increasing values of \(N\), represented from lighter to darker colors under  correlated and local amplitude damping reservoirs. (b) \(\mathcal{C}_E/N\) in the energy basis of the battery Hamiltonian  (vertical axis) vs  time (horizontal axis). Dashed and solid lines denote local  and spatial correlated baths respectively. Initially, the state is prepared as a product state, \(|-\rangle^{\otimes N}\). Other parameters are \(\ \gamma_{12}=0.01e^{i\pi/3}\), \(D=0.2\), \(J=1.2\)  and \(h=1.0\) for the battery Hamiltonian. All the axes are dimensionless.}
    \label{fig:loc_bat_adc}
\end{figure}
\begin{figure}
    \centering
    \includegraphics[width=\linewidth]{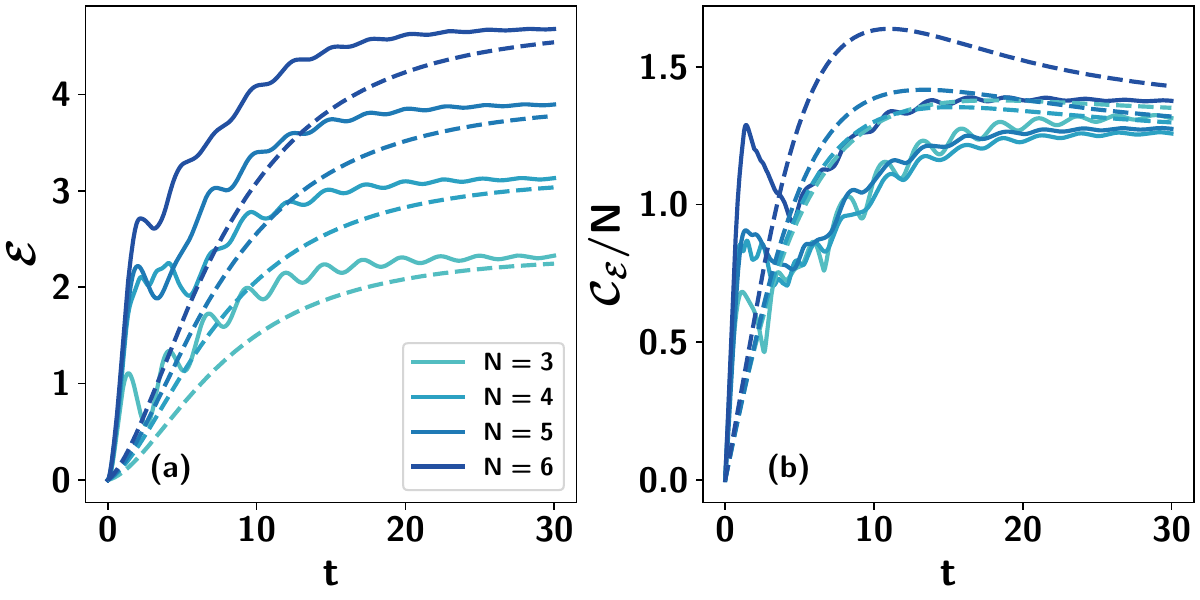}
    \caption{When the initial state is chosen to be the ground state of the battery Hamiltonian with \(h=1.3\), which turns out to be entangled, the dynamics of the ergotropy and the coherence per site (vertical axis) are plotted with respect to  time (horizontal axis). All other specifications are the same as in Fig. \ref{fig:loc_bat_adc}.
    All the axes are dimensionless.}
    \label{fig:cor_bat_adc}
\end{figure}

\subsection {Behavior of ergotropy: Local vs correlated amplitude damping noise}

As we have analyzed in the pervious cases, we again consider the initial battery system to be a product state (with \(h \gg J\)), and an entangled state.

\emph{Initial product battery state.} 
Preparing the states as \(|-\rangle^{\otimes N}\), the evolved state in arbitrary time \(t\) under local noise takes the form \(\rho_B(t)=\rho(t)^{\otimes N}\) where the \(\rho(t)\) is 
\begin{equation}
   \rho(t)= \left [
\begin{array}{cc}
 1-\frac{e^{\text{-$\gamma $} t}}{2} & -\frac{e^{-\text{$\gamma $} t/2}}{2} \\
 -\frac{e^{-\text{$\gamma $} t/2}}{2} & \frac{e^{\text{$-\gamma $} t}}{2} \\
\end{array}
\right ],
\end{equation}
with \(\gamma_{ii} = \gamma\) as chosen before.
One can note that although the resulting state is not entangled, it is capable of altering the population of the energy levels, which, in turn, modifies the ergotropy of the state. As a consequence, a non-zero steady-state ergotropy can be obtained in the battery. Moreover, since the noise influences the level populations, it also induces coherence in the energy basis. Unlike the dephasing channel, the amplitude damping channel generates coherence in the energy basis, thereby enabling nonvanishing ergotropy even under the local noise. Nevertheless, the gain in ergotropy in presence of correlated amplitude damping noise remains evident, particularly, in the transient regime (see Fig.~\ref{fig:loc_bat_adc}). Specifically, we observe that the ergotropy as well as coherence in the correlated case is enhanced during the transient regime and exhibits oscillations before saturating at the steady-state value, which increases with the system-size \(N\). Importantly, the steady-state saturation ergotropy coincides with that of the local noise case, since after a critical time, the local component of the correlated noise dominates the dynamics. In fact, while the steady-state ergotropy increases with the system-size \(N\), the steady-state coherence decreases with \(N\). Thus, coherence in the energy basis cannot be considered as a universal resource, although it provides a consistent explanation in the dephasing noise case. Further, the correlated noise may generate entanglement between the parties in the transient time while such correlation vanishes at the steady state.

\emph{Role of the initially entangled state.} We now turn to the role of initially entangled states, i.e., when the interaction between the subsystems of the battery is switched on. 
The behavior of ergotropy does not alter in the presence of initial entanglement within the battery, underscoring its beneficial role across a wide range of reservoirs and initial states (see Fig.~\ref{fig:cor_bat_adc}). Importantly, the entangled initial state consistently yields higher ergotropy compared to the product state, both in the transient regime and in the long-time steady state, thereby confirming the advantage of initial correlations in energy extraction. Furthermore, our findings establish that amplitude damping noise invariably produces greater ergotropy than the dephasing noise, regardless of whether the reservoirs are correlated or local. This demonstrates the inherent superiority of amplitude damping environments and emphasizes the combined impact of entanglement and noise structure in improving the battery performance.

\begin{figure}
    \centering
    \includegraphics[width=\linewidth]{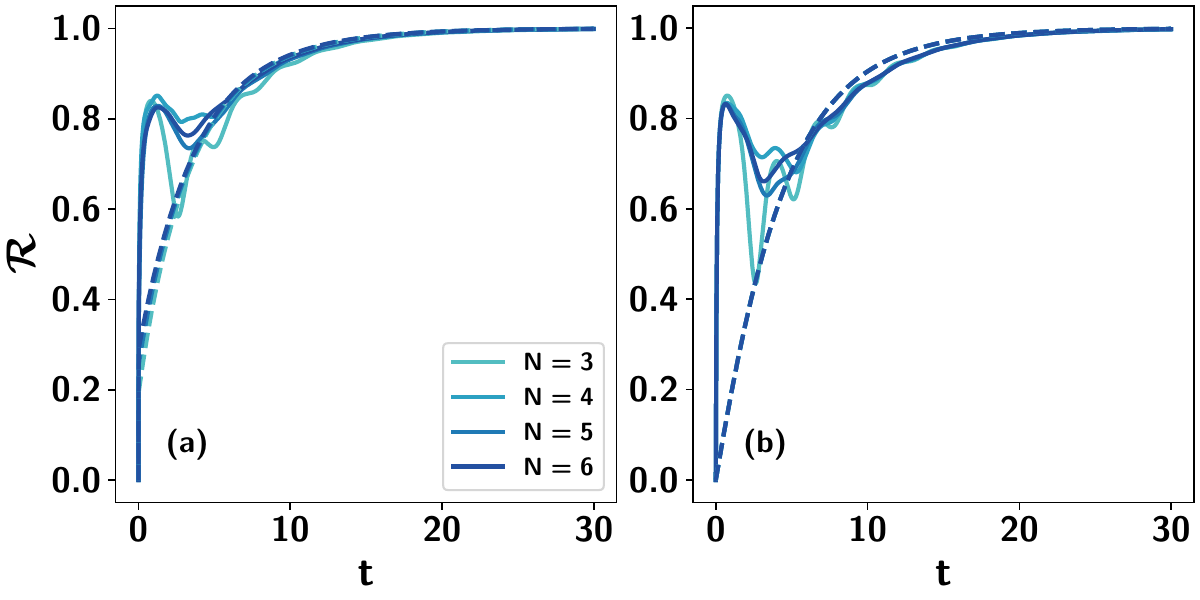}
    \caption{\(\mathcal{R}\)  (ordinate) vs time (abscissa) for (a) unentangled and (b) entangled initial states when they are in contact with correlated (solid) and local (dashed) amplitude damping  noise. As in Fig. \ref{fig:loc_bat_adc}, the advantage of correlated reservoirs is evident in the transient regime irrespective of the initial states.  Further, \(\mathcal{R}\) has no effect on system-size for local noise. All other parameters are the same as in Fig. ~\ref{fig:cor_bat_adc}. All the axes are dimensionless.}
    \label{fig:ratio_ergo_energy}
\end{figure}

\emph{Fraction of extractable energy.} In contrast to the dephasing noise, under amplitude damping noise, the fraction \(\mathcal{R}\) increases with time and eventually saturates to a finite value, which is the maximum ergotropy achievable, independent of whether the initial state is entangled or a product state (see Fig.~\ref{fig:ratio_ergo_energy}). Again, from small to moderate time periods, \(\mathcal{R}\) is consistently higher for correlated reservoirs compared to local noise. In the steady state, however, the fraction reaches unity, indicating that all stored energy becomes extractable, and that the performance of the battery against correlated and local noise coincides.

Our results, both for amplitude and phase damping noises reveal that the reservoir's correlation can be used to enhance the efficiencies in the quantum devices like quantum thermal machines, quantum sensors, etc for small time which may not remain so for large time scale. However, such benefits are aligned with the experimental set-up as the typical device's performances are extracted only during finite time scale.

\section{Role of effective Long-range interactions in dynamics of the battery}
\label{sec:LRint}

Until now, we observed the beneficial effect of the correlated reservoirs on the extractable work of the battery, where an effective interaction is introduced between the nearest-neighbor reservoirs. It is now tempting to find out that if one extends correlation between reservoirs beyond nearest neighbors, whether we can obtain higher efficiency from quantum devices than the one reported until now. To answer it, an all-to-all interaction in the effective dephasing and amplitude damping Hamiltonian in Eqs. (\ref{eq:corrdeph}) and (\ref{eq:adccorr}) is incorporated. Hence, the corresponding effective interaction Hamiltonians for the dephasing and amplitude damping noises respectively become 
\begin{eqnarray}
    \mathcal{H}_{z} (t) = \sum_{j<k=1}^{N}\mathcal{J}^{z} \hat{\sigma}_j^z\hat{\sigma}_{k}^z,
    \end{eqnarray}
\begin{eqnarray}
    \mathcal{H}_{xy} = \sum_{j<k=1}^N \mathcal{J} \hat \sigma_{j}^{+} \hat \sigma_{k}^{-} + \mathcal{J^{*}} \hat \sigma_{j}^{-} \hat \sigma_{k}^{+}
    \label{eq:lr_ham}
\end{eqnarray}
In this framework, the \(\hat\Gamma^z\) matrix is expressed as
\begin{equation}
    \hat\Gamma^z = \begin{bmatrix}
\gamma & \gamma_{12} & \gamma_{13} & \cdots & \gamma_{1N-1} & \gamma_{1N}^* \\
\gamma_{12}^* & \gamma & \gamma_{23} & \ddots & & \gamma_{2N} \\
\gamma_{13}^* & \gamma_{23}^* & \gamma & \ddots & \ddots & \vdots \\
\vdots & \ddots & \ddots & \ddots & \gamma_{N-2N-1} & \gamma_{N-2N} \\
\gamma_{1N-1}^* & & \ddots & \gamma_{N-1N-1}^* & \gamma & \gamma_{N-1N} \\
\gamma_{1N} & \gamma_{2N} & \cdots & 0 & \gamma_{N-1N}^* & \gamma
\end{bmatrix}.
\label{eq::gama_matrix2}
\end{equation}
In our study, we consider all \(\gamma_{i\neq j} = \beta*e^{i\phi} = \gamma_{12}\), i.e., the interaction strength between all the reserviors are same. Now, the eigenspectrum of the matrix \(\hat{\Gamma}^z\) is not completely solvable, but the bound on the eigenvalues can be given, which ensures the dynamics to be a CPTP map. Specifically, the dynamical map is a valid CPTP operation if the following condition holds:
\begin{eqnarray}
\gamma \geq (N-1)|\gamma_{ij}| \quad \forall i\ne j,
\end{eqnarray}
where \(N\) is the number of spins in the system.
In order to calibrate the effect of all-to-all interaction between the environments, we choose the parameters of the reservoirs accordingly for a given system-size.

\begin{figure}
    \centering
    \includegraphics[width=\linewidth]{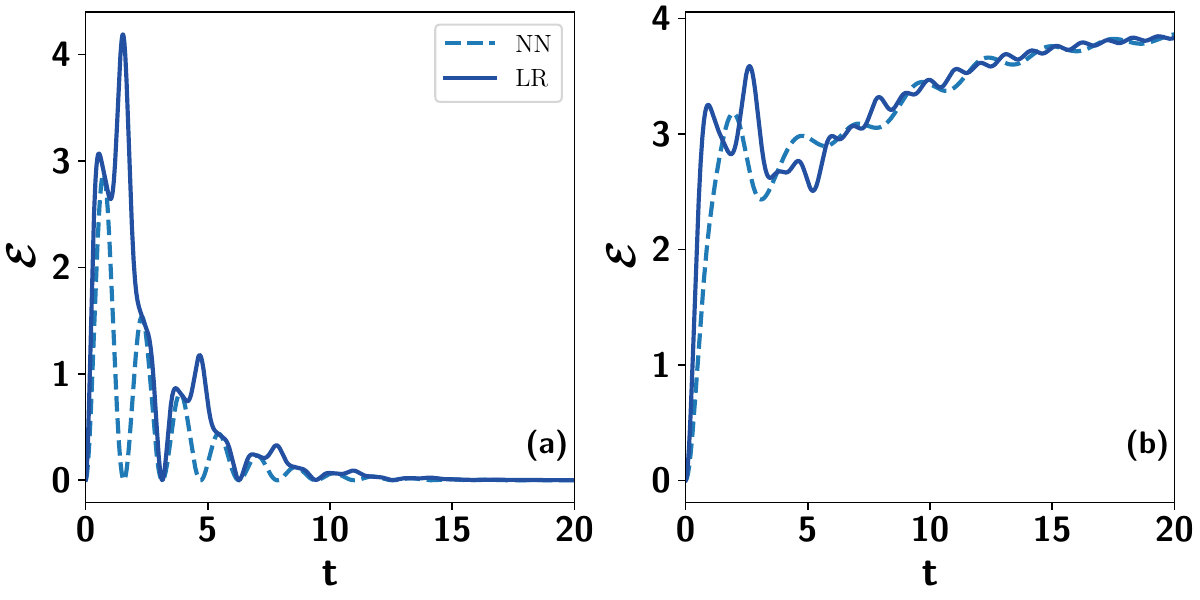}
    \caption{Effects of long-range interactions among reservoirs. Ergotropy (ordinate) against time (abscissa) for (a) dephasing and (b) amplitude damping spatially correlated noise. Solid and dashed lines  represent effective long-range (LR) and nearest-neighbor (NN) interactions in GKSL master equation respectively. The initial state is the ground state of the battery Hamiltonian with \(J'=0.0\), i.e., \(|-\rangle^{\otimes N}\). Here \(N=6\), \(\gamma=0.2\), \(\gamma_{ij}=0.01e^{i\pi/3}\), \(D=0.2\),  \(J=1.2\), and \(h=1.3\). All axes are dimensionless.}
    \label{fig:de_ad_all}
\end{figure}
The overall trends of ergotropy for the initial state, \(|-\rangle^{\otimes N}\) are the same, both for the correlated dephasing and amplitude damping noises. Comparing nearest-neighbor interactions with the long range (LR) ones, we observe that the ergotropy is higher for the later case than the NN one for short time period, thereby highlighting the significance of correlations among reservoirs on quantum devices (see Fig.~\ref{fig:de_ad_all}).  However, as expected,  the influence of NN and LR interactions of the reservoirs on the extractable work of the battery does not differ  for a large time, because they coincide with the local noise. However, in the case of amplitude damping noise, the saturation happens slowly and the ergotropy is higher in this case, compared to the dephasing correlated noise, as found in the NN case also.

\section{Conclusion}
\label{sec:conclu}


At the end of the last century, it was recognized that exploiting quantum mechanical principles could enable the construction of more efficient and powerful devices than the existing classical ones. A prominent example is quantum thermal machines, which include quantum analogs of energy extraction and storage devices, known as quantum batteries. Beyond their technological promise, these devices also play a crucial role in deepening our understanding of thermodynamic principles at the microscopic scale.

A key resource for advancing quantum technologies lies in the intrinsic quantum properties of a system, such as entanglement and coherence, which, however, typically diminish under environmental interactions. In this work, we emphasized on two distinct advantages of these quantum features: one stemming from correlations within the environment itself, and the other from coherence and entanglement embedded in the battery. We showed that correlations between reservoirs can enable finite extractable work, or ergotropy, from a quantum battery in the transient regime, even when the battery cells are initialized in a product state, something impossible under purely local baths. To demonstrate this, we employed spatially correlated dephasing and amplitude-damping channels with a moderate number of battery cells. Our findings revealed that coherence, computed in the energy eigenbasis of the battery Hamiltonian, serves as the key resource driving energy extraction, especially in the dephasing case. Furthermore, we observed that long-range correlations among reservoirs can enhance the performance of the battery compared to the nearest-neighbor correlations for a small time. On the other hand, when the initial state of the battery is prepared in an entangled state, we showed that the fraction of extractable energy remains finite under both noise models, even at the steady state, and surpasses that obtained with local noise in the transient time. Notably, while a product initial state under dephasing noise yields nonvanishing ergotropy, this advantage persists only in the transient regime.

These insights resonate with the growing body of research on reservoir engineering, which seeks to harness environmental effects rather than merely mitigate them, for the purpose of building more efficient quantum devices under decoherence. Our findings contribute meaningfully to this emerging direction by demonstrating how correlations within the environment, alongside initial-state quantum resources, can enhance device performance. In particular, this work offers valuable guidance for designing robust quantum thermal machines that can operate effectively even in the presence of environmental noise.

\acknowledgements
 We acknowledge the use of \href{https://github.com/titaschanda/QIClib}{QIClib} -- a modern C++ library for general purpose quantum information processing and quantum computing (\url{https://titaschanda.github.io/QIClib}), and the cluster computing facility at the Harish-Chandra Research Institute. This research was supported in part by the INFOSYS scholarship for senior students. We acknowledge support from the project entitled ``Technology Vertical - Quantum Communication'' under the National Quantum Mission of the Department of Science and Technology (DST)  (Sanction Order No. DST/QTC/NQM/QComm/$2024/2$ (G)).

\appendix

\section {Emergence of Ising interaction in the case of dephasing noise}
\label{sec:appendixA}

Consider a quantum battery (\(H_B\)) made up of \(N\) spins, each coupled to its own reservoir (\(H_E\)), subject to local and correlated noises (the Hamiltonian representing system-environment interaction,  \(H_{BE}\)). The GKSL master equation as given in the main text reads as
\begin{eqnarray}
\nonumber \dot\rho_B(t) &=& -\frac{1}{\hbar^2}\int_{0}^{t} d\tau [\text{Tr}_{E}[H_{BE}(t), \\ && [H_{BE}(\tau),\rho_B(t)\otimes\rho_E]]].
\label{eq:master_eq}
\end{eqnarray}
The above equation assumes the initial state to be a product state of the system and the environment.
Employing the form of system-environment interaction $H_{BE}^D$ for the dephasing noise from the Eq. (\ref{eq:system_bath_coupling}) and taking the trace over the environment, the expression becomes 
\begin{eqnarray}
\nonumber \dot\rho_B (t) &=& -\frac{1}{\hbar^2}\sum_{r,s}\int_{0}^t d\tau [\Gamma_{rs}(t-\tau)(\hat \sigma_{r}^z \hat \sigma_{s}^z \rho_B - \hat \sigma_{s}^z \rho_B \hat \sigma_{r}^z) + \\ && \Gamma_{sr}(\tau-t)(\rho_B \hat \sigma_s^z \hat \sigma_r^z - \hat \sigma_r^z \rho_B \hat \sigma_s^z)],
\end{eqnarray}
where the two point correlation function \(\Gamma_{rs}\) as given in the main text, is defined as
\begin{eqnarray}
\Gamma_{rs} (t-\tau) \equiv \langle G_r (t) G_s(\tau) \rangle 
\end{eqnarray}
with \(G_r(t) = e^{iH_Et/\hbar}G_r(0)e^{-iH_Et/\hbar}\) evaluated in the interaction picture and \(\langle \mathcal{Q} \rangle \equiv \text{Tr}_E(\rho_E \mathcal{Q})\), \(G_r\) being an operator acting on the environment.
After scaling as \((t - \tau) \rightarrow \tau\) and rearranging the terms, the final expression is given as \cite{Zou2024}
\begin{eqnarray}
\nonumber \dot\rho_B (t) &=& -\frac{1}{2\hbar^2}\sum_{r,s}\int_0^t d\tau [\Gamma_{rs}(\tau) - \Gamma_{rs}(-\tau)][\hat \sigma_{r}^z \hat \sigma_{s}^z, \rho_B] + \\ && \frac{1}{\hbar^2}\sum_{r,s}\int_{-t}^t d\tau  \Gamma_{rs}(\tau)\mathcal{K}_{rs}^z[\rho_B],
\end{eqnarray}
Hence, we obtain the GKSL master equation as in Eq. (\ref{eq:mastereqn}) for the correlated dephasing noise, 
\begin{eqnarray}
{\dot\rho_B(t)} &=& -i [\mathcal{H}_{z} (t), \rho_B(t)] + \sum_{r,s = 1}^{N} \gamma_{rs}^z\mathcal{K}_{rs}^{z} [\rho_B(t)],
\end{eqnarray}
where the coherent interaction parameter, dissipation term and the dissipation strength, respectively, with the Lindblad operator \(\{\hat \sigma^z\}\), become
\begin{equation}
    \mathcal{J}^z(t) = \frac{1}{2i\hbar^2}\sum_{i\neq j}\int_0^t d\tau [\Gamma_{rs}(\tau) - \Gamma_{ij}(-\tau)],
\end{equation}
\begin{equation}
   \mathcal{K}_{rs}^z[\rho] \equiv \hat \sigma_s^z \rho \hat \sigma_r^z - \{\hat \sigma_r^z \hat \sigma_s^z,\rho\}/2, 
\end{equation}
\begin{equation}
    \gamma_{rs}^z (t) = \frac{1}{\hbar^2} \int_{-t}^{t} d\tau \Gamma_{rs}(\tau)
\end{equation}
In general, the parameters \(\mathcal{J}\) and \(\gamma_{rs}^z\) are time-dependent, corresponding to a particular form of the correlation function, as is evident from the above expressions. However, in our case, we have considered the Markovian limit, where these can be taken as time-independent and constants.

\section {Emergence of DM interaction in the case of amplitude damping noise}
\label{sec:appendixB}

Considering the amplitude damping noise, the Hamiltonian \(H_{BE}^{A}\) taken from Eq. (\ref{eq:system_bath_coupling}), in the interaction picture, becomes
\begin{eqnarray}
H_{BE}^{A} (t)=\sum_{j=1}^{N} e^{i h t}\hat{\sigma}_{j}^+\otimes G_j+e^{-i h t}\hat{\sigma}_{j}^-\otimes G_j^\dagger,
\end{eqnarray}
In order to obtain the GKSL master equation for the above case, with \(H_{BE}^{A}\) being the interaction Hamiltonian between the system and the environment, and  considering the weak coupling limit, the evolution of the battery is given as  \cite{Zou2024,driessen2025}
\begin{equation}
\begin{split}
 \dot\rho_B (t) = &-\frac{1}{\hbar^2}\sum_{r,s}\int_0^t d\tau [\Gamma_{rs}(\tau)(e^{ih\tau} (\hat \sigma_{r}^+ \hat \sigma_{s}^- \rho_B - \hat \sigma_{s}^- \rho_B \hat \sigma_{r}^+ ) \\ 
&+ e^{-ih\tau}(\hat \sigma_{r}^- \hat \sigma_{s}^+ \rho_B
- \hat \sigma_{s}^+ \rho_B \hat \sigma_{s}^-)) \\
&+ \Gamma_{rs}(-\tau)(e^{-ih\tau}(\rho_B \hat \sigma_{r}^+ \hat \sigma_{s}^- - \hat \sigma_{s}^- \rho_B \sigma_{r}^+ ) \\
&+ e^{ih\tau} (\rho_B \hat \sigma_{r}^- \hat \sigma_{s}^+ - \hat \sigma_{s}^+ \rho_B\hat \sigma_{r}^-)],
\end{split}
\end{equation}
After rearranging the terms, we obtain the GKSL master equation, given in Eq. (\ref{eq:mastereqn_adc}) as
\begin{eqnarray}
{\dot\rho_B(t)} &=& -i[\mathcal{H}_{xy}, \rho_B(t)] + \sum_{r,s =1}^N \mathcal{L}_{rs}(t) \rho_B(t),
\end{eqnarray}
where the Lindbladian, with Lindblad operators as \(\{\hat \sigma^+, \hat \sigma^-\}\) becomes
\begin{eqnarray}
\nonumber \mathcal{L}_{rs} (t) \rho_B (t) = \gamma_{rs}^{\uparrow}(t)[\hat \sigma_{s}^+\rho_B\hat \sigma_r^- - \frac{1}{2} \{\hat \sigma_r^- \hat \sigma_s^+,\rho_B \}] && \\ + 
\gamma_{rs}^{\downarrow}(t)[\hat \sigma_{s}^-\rho_B \hat \sigma_r^+ - \frac{1}{2} \{\hat \sigma_r^+ \hat \sigma_s^-,\rho_B\}].
\end{eqnarray}
Here, \(\gamma^{\downarrow}_{rs} (t)\) and \(\gamma^{\uparrow}_{rs} (t)\) represent the decay and excitation rates respectively for local (\(r=s\)) and correlated (\(r \neq s\)) noises. In our case, we have considered zero-temperature limit, where no excitations happen, i.e., \(\gamma^{\uparrow}_{rs}(t) = 0\). Moreover, considering Markovian limit, we get \(\gamma^{\downarrow}_{rs}(t) = \gamma_{rs}(t) = \gamma_{rs}\) and we achieve the final equation as
\begin{equation}
\begin{split}
\dot\rho_B(t) &= -i[\mathcal{H}_{xy}, \rho_B(t)] \\ 
&+ \sum_{r,s =1}^N\gamma_{rs}(t)[\hat \sigma_{s}^-\rho_B \hat \sigma_r^+ - \frac{1}{2} \{\hat \sigma_r^+ \hat \sigma_s^-,\rho_B\}],
\end{split}
\end{equation}
which we use to derive the dynamics of the battery state and compute the ergotropy.

\bibliography{ref.bib}
\end{document}